# Exploring Morphology-Activity Relationships: *Ab Initio* Wulff Construction for RuO$_2$ Nanoparticles under Oxidizing Conditions

Tongyu Wang[a], Jelena Jelic[a], Dirk Rosenthal[b], and Karsten Reuter*[a]

[a] Chair for Theoretical Chemistry and Catalysis Research Center, Technical University Munich, Lichtenbergstr. 4, D-85747, Germany

[b] Fritz Haber Institute of the Max Planck Society, Faradayweg 4-6, D-14195, Berlin, Germany

## Abstract

We present a density-functional theory based Wulff construction of the equilibrium shape of RuO$_2$ particles in an oxygen environment. The obtained intricate variations of the crystal habit with the oxygen chemical potential allow for a detailed discussion of the dependence on the oxidizing pretreatment observed in recent powder catalyst studies. The analysis specifically indicates an incomplete particle shape equilibration in previously employed low temperature calcination. Equilibrated particles could be active CO oxidation catalysts with long-term stability in oxidizing feed and then represent an interesting alternative to the previously suggested core-shell concept.

# Introduction

Systematic studies on polycrystalline powders are an appealing link between detailed Surface Science work on single crystals and the supported nanoparticles in real catalysis. They offer a general route to the exploration of important factors like particle size, shape and composition without the additional complexity introduced by a support material. At present such structure-morphology-activity relationships are only scarcely established though. Not least, this is due to experimental difficulties in determining the crystal habit, i.e. particle shapes and (atomic-scale) surface structure of the exposed facets, under reaction conditions. With ambient gas phases severely challenging the imaging and analytical capabilities of e.g. electron microscopes [1], *ex situ* (either as prepared or *post mortem*) studies still mostly provide the only access. This, however, neglects the attested possibly strong dynamical evolution of operating catalysts and the significant impact this can have on the catalytic function [2-4]. Even just the effect of catalyst pretreatments on surface phases and particle shapes is often not systematically explored, which determines the induction behaviour during the dynamical evolution and which by itself may contain valuable information about the active sites involved.

In this situation independent information from first-principles based theory can be an enormous asset. To this end and complementing the experimental advances towards an *in situ* characterization, recent years have brought significant progress towards the consideration of finite gas phases in corresponding electronic structure calculations [5,6]. In particular, density-functional theory (DFT) based *ab initio* thermodynamics approaches [7-9] are now readily applied to obtain the surface structure and composition of single-crystal model catalysts in contact with a multi-component gas-phase environment. Combining the corresponding surface free energies within a Wulff construction then yields particle shapes in explicit dependence on temperature and reactant partial pressures [10-12].

We here further explore this concept to address the largely different initial activities towards the CO oxidation reaction obtained in recent studies on polycrystalline $RuO_2$ powder catalysts by Narkhede *et al*. [13] and Rosenthal *et al*. [14-15]. They have been related to different calcination temperatures during catalyst preparation, yielding distinctly different crystal habits [13,14]. Understanding these reactivity differences in terms of the different facets and surface terminations exposed then offers an interesting new perspective on the controversially discussed active phases of Ru catalysts for this reaction. These catalysts exhibit dramatic activity variations as a function of reactant pressure, which are by now generally assigned to a change in the oxidation state [16-18]. However, despite significant efforts, not even consensus has been reached regarding the detailed active state of the close-packed Ru(0001) facet [19-21], i.e. whether it corresponds to the pristine metal with some coverage of oxygen adsorbates [22], a more or less bulk-like film of $RuO_2(110)$ [13,14,23,24], or some intermediate (possibly badly defined) oxygen-enriched surface fringe involving sub-surface oxygen or a transient surface oxide [14,15,25-28]. With $RuO_2(110)$ itself clearly exhibiting high activity, this situation is further complicated by an enigmatic deactivation of supported Ru catalysts under oxidizing conditions [29,30], which was equally observed for polycrystalline $RuO_2$ particles [31].

For the latter particles, this deactivation has been assigned to a microfacetting of the lateral $RuO_2(110)$ facets into an inactive $RuO_2(100)$-c(2x2) phase [31]. On the basis of our first-principles Wulff particle shapes we argue though that the presence of these lateral {110} facets (and the concomitant deactivation) could merely result from an incomplete particle shape equilibration at the employed low calcination temperature. Different oxidizing pretreatments are indeed found to critically affect the crystal habit, in particular changing the cross section from quadratic to octagonal, as well as the share and type of facets exposed. Correlating this information with the reported activity data points to an important role of the hitherto not much studied {101} and {111} apical facets, and specifically to their oxidation state under steady-state operation in net oxidizing or stoichiometric feed.

## Results and Discussion

The objective of our study are the different $RuO_2$ particle shapes obtained after different calcination procedures, as well as the different initial activities under net-oxidizing feed, where the $RuO_2$ bulk structure was shown to be intact [13-15]. Narkhede *et al.* [13] calcined in a 10% $O_2$/Ne mixture at 573K, while Rosenthal *et al.* [14-15] employed pure oxygen at 1073K. We address these findings with a first-principles atomistic thermodynamics Wulff construction for $RuO_2$ particles in a pure oxygen environment [10]. Due to the tetragonal symmetry of bulk rutile, this requires to determine the surface free energies of the five inequivalent low-index surfaces: (100), (110), (111), (101) and (001). For each orientation we consider all possible (1x1)-terminations, which result from cutting the crystal stacking at different planes. For all five orientations this yields an O-poor, a stoichiometric, and an O-rich termination, while for the (111) orientation a second even more O-rich termination (coined super-O-rich) is also possible. In this super-O-rich termination the fivefold-coordinated surface $Ru_{cus}$ atom of the O-rich termination is capped by an on-top O atom at a bond distance of 1.67 Å, i.e. a situation that occurs equivalently already in the O-rich termination of the most studied $RuO_2$(110) surface [7]. Unfortunately, the geometric structure of the c(2x2)-$RuO_2$(100) reconstruction is as yet elusive, which is why we cannot explicitly compute its surface free energy. Instead, we will derive bounds for its value on the basis of the obtained Wulff shapes.

In equilibrium with the surrounding $O_2$ gas phase the most stable termination will minimize the surface free energy [7,8]

$$\gamma(T,p) = \frac{1}{A}\left[G(T,p,N_{Ru},N_O) - N_{Ru}\mu_{Ru}(T,p) - N_O\mu_O(T,p)\right]. \quad (1)$$

Here, $A$ is the surface area, $G$, $\mu_{Ru}$, $\mu_O$ are the Gibbs free energy of the solid with the surface, the Ru and O chemical potentials, respectively, and $N_{Ru}$ and $N_O$ are the number of Ru and O atoms in the solid with the surface. For not too low temperatures and sufficiently large particles we can assume the surface to be equilibrated with bulk $RuO_2$, which allows to eliminate $\mu_{Ru}$ in eq. (1) in favour of the Gibbs free energy of bulk $RuO_2$ [10]. Neglecting vibrational free energy contributions in the difference of bulk and surface Gibbs free energy, the remaining difference in total energies entering eq. (1) is computed within semi-local density-functional theory [7,8].

The effect of the surrounding gas phase enters through the free energy contribution $\Delta\mu_{O2}(T,p)$ to the oxygen chemical potential $\mu_O(T,p) = \frac{1}{2} E_{O2} + \frac{1}{2} \Delta\mu_{O2}(T,p)$, where $E_{O2}$ is the total energy of an isolated $O_2$ molecule (including zero-point energy). Figure 1 illustrates the resulting dependence on the oxygen chemical potential for the surface free energies of the four terminations considered for the $RuO_2$(111) surface, where ideal gas laws are employed to directly convert this dependence into a temperature scale at fixed $O_2$ pressure of 1 atm. Additionally shown are approximate, but well-defined limits for the chemical potential range of interest [7,10]: The upper O-rich boundary roughly denotes the onset of oxygen condensation on the sample and the lower O-poor boundary denotes the stability limit of bulk $RuO_2$ against decomposition into Ru metal, $O_2$, or volatile $RuO_x$ species. Apart from the stoichiometric termination all other terminations exhibit an explicit dependence on the O chemical potential, with e.g. the O deficient O-poor termination becoming more favorable at higher temperatures. Through this dependence we obtain a change in the most favorable termination over the range of chemical potentials: At lower chemical potentials (higher temperatures) the O-rich termination results as most stable, whereas at higher chemical potentials (lower temperatures) this is the super-O-rich termination.

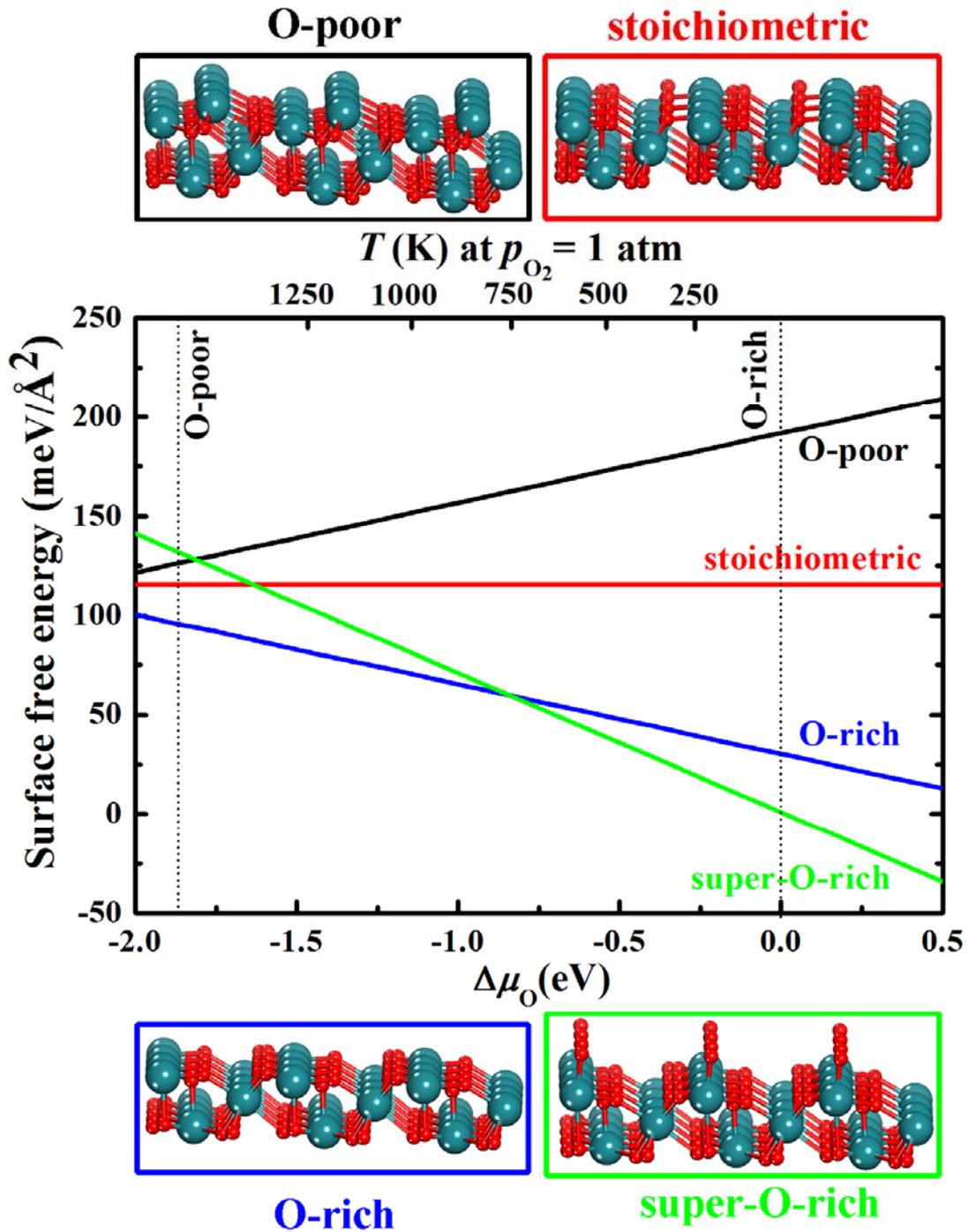

Figure 1: Surface free energies and surface geometries of the four possible (1x1) terminations of RuO$_2$(111). The dependence on the O chemical potential is converted into a temperature scale at p$_{O2}$ = 1 atm in the upper x-axis. The vertical dotted lines represent the ΔµO range considered in this study, with the O-poor limit denoting the stability limit of bulk RuO$_2$ (see text).

Figure 2 compiles the equivalent results obtained for all five low-index orientations. For each orientation we now directly show the minimized surface free energy, where a change of slope reflects a change in the most stable termination. We obtain such a change of the most stable termination for all five orientations. For $RuO_2(111)$ this is as discussed a transition from O-rich to super-O-rich. In all other cases this is a transition from stoichiometric to O-rich towards higher chemical potentials. This generalizes the findings reported before for the $RuO_2(110)$ surface [7], namely the prediction of O-enriched "polar" terminations at higher chemical potentials e.g. for near ambient $O_2$ pressures in the temperature range below ~600 K, cf. Fig. 2. The existence of such terminations in turn reveals the shortcomings of the frequent notion to identify "most stable" crystal faces on the basis of the surface energy of the stoichiometric termination [32,33]. In fact, the (environment-independent) computed surface energies of these terminations do reflect the generally expected ordering, with the $RuO_2(110)$ surface exhibiting the lowest value. However, as we will exemplify below this does not necessarily mean that {110} facets have a large share of the crystal surface for gas-phase conditions corresponding to higher chemical potentials where other terminations might get stabilized.

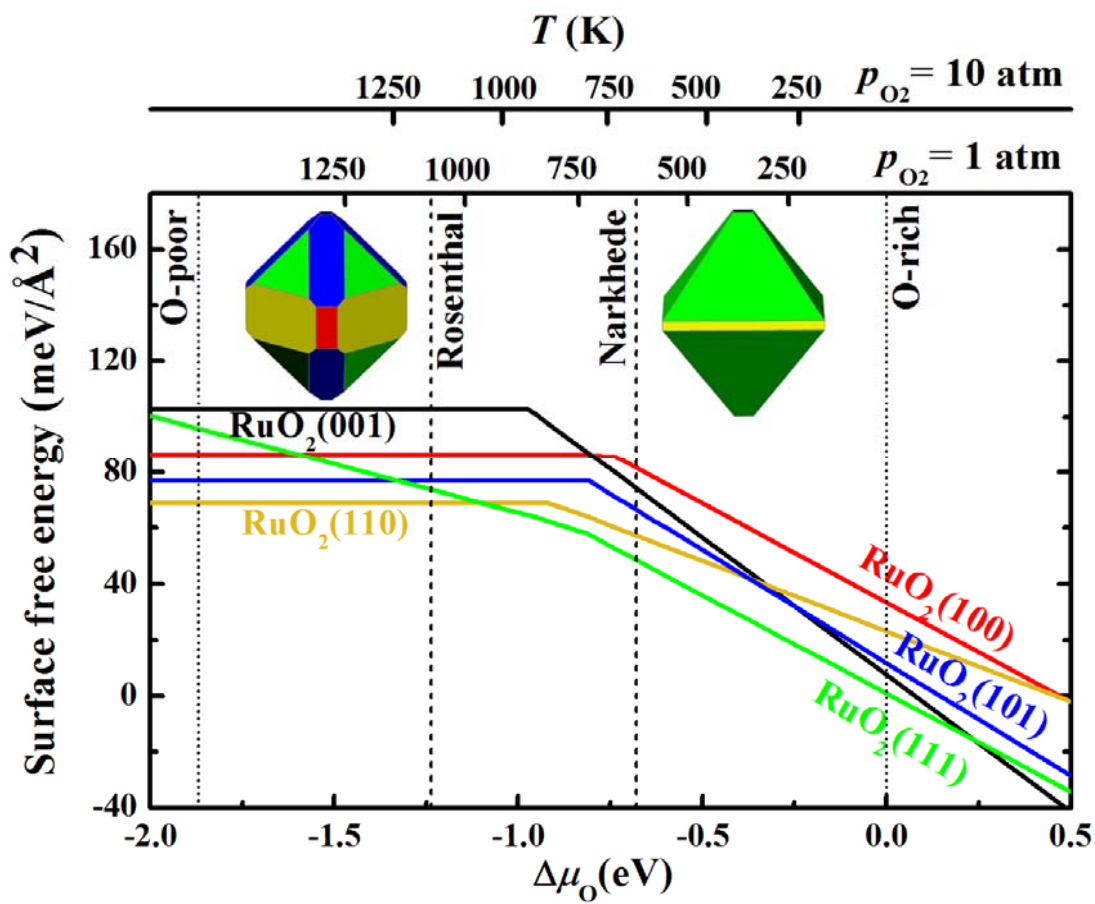

Figure 2: Same as Fig. 1, but showing the minimized surface free energies for the five low-index RuO2 surfaces. Additionally indicated by vertical dotted lines are the O chemical potentials corresponding to the calcination procedures employed in the studies by Narkhede et al. and Rosenthal et al., (0.1atm, 573K) and (1atm, 1073K), respectively [13-15]. The insets show the corresponding particle shapes obtained from the constrained Wulff construction (see text), where the colors of the different facets match those of the corresponding surface free energy lines.

With these results for the different orientations we can proceed to evaluate a constrained Wulff shape that minimizes the total surface free energy for given volume [34]. With "constrained" we hereby indicate that the input to this construction is restricted to the surface free energies of the computed (1x1)-terminations. Rather than determining the real equilibrium $RuO_2$ crystal shape, the purpose of the construction is thus more to compare the relative energies of the different (1x1)-terminated surface orientations and thereby establish a basis to discuss the effect of surface reconstructions on the crystal habit. Since the surface free energies depend on the oxygen chemical potential, the constrained Wulff shape can vary with the gas-phase conditions and in the present context specifically with the applied calcination procedure.

As shown in Fig. 2 notably different shapes are indeed obtained for the two procedures employed in the powder catalyst studies. Apart from the similarity of a small {001} plateau capping the (symmetry-equivalent) apical sides, these differences extend to both the exposed lateral and apical facets: At the (0.1 atm, 573K) employed by Narkhede *et al.* [13] the Wulff particle has a quadratic cross-section and only exhibits {111} apical and {110} lateral facets. In contrast, at the (1atm, 1073K) employed by Rosenthal *et al.* [14-15] the Wulff particle additionally features {101} apical and {100} lateral facets, with the cross-section concomitantly changed to irregular octagonal. The primary reason behind these rather significant differences in the crystal habit is the difference in terminations stabilized at the corresponding O chemical potentials. At the conditions of the higher temperature calcination $\Delta\mu_O$ is low enough to stabilize the stoichiometric terminations at all low-index orientations apart from the {111} facets, cf. Fig. 2. On the contrary, the higher chemical potential corresponding to the conditions of the lower temperature calcination falls already into the stability range of the O-rich terminations (or super-O-rich termination in case of the {111} facets). The concomitant pronounced lowering of the (111) surface free energy preferentially stabilizes this apical facet, and the changed ratio of the surface free energies of the (110) and (100) orientations leads to the cross-section change: For $1/\sqrt{2} < \gamma(100)/\gamma(110) < \sqrt{2}$, both facets are exhibited at irregular octagonal cross-section. For $\gamma(100)/\gamma(110) < 1/\sqrt{2}$ or $\gamma(100)/\gamma(110) > \sqrt{2}$, the cross-section is quadratic, with only {100} or {110} facets exposed, respectively.

Overall the obtained particle shapes are quite consistent with the electron microscopy images reported in the two powder catalyst studies [13,14]. This holds in particular for the cross-sections and type of apical and lateral facets exposed. In both cases, the experimental shapes are much more columnar with a larger share of the lateral facets though. Similar habits as the ones of the Rosenthal crystals have been reported in earlier growth studies that even involved higher calcination temperatures [35,36]. For the Rosenthal crystals this renders kinetic limitations as explanation for the difference to our computed Wulff shape unlikely. Instead, we attribute this difference to the previously reported restructuring of {110} facets into c(2x2)-{100} microfacets [31] that is not contained in our model. Neglecting the actual cost of the microfacetting itself, such a restructuring would equalize the (110) and (100) surface free energies at a lower value. The concomitant change of the $\gamma(100)/\gamma(110)$ ratio towards one would make the cross-section even more octagonal, and the surface free energy lowering of the lateral facets would elongate the crystal shape. This is illustrated in Fig. 3b, where the surface free energy of all lateral facets has at least been reduced to the value of the stoichiometric (110) termination. In reality, the surface free energy of the c(2x2)-(100) reconstruction must be (much) lower than this though, otherwise there would be no thermodynamic driving force for the microfacetting of the {110} facets. As shown in Fig. 3c, there is, however, a lower bound for this value, too, since a too strong lowering of the surface free energy of the lateral facets would eliminate the {111} apical facets, which are clearly seen in experiment. At the $\Delta\mu_O$ = -1.24 eV corresponding to the Rosenthal calcination, this is obtained, when lowering the surface free energy of the {100} facets from 86 meV/Å$^2$ for the stoichiometric (1x1)-termination by 69% to 27 meV/Å$^2$.

In the case of the Narkhede experiments, the reported more columnar, needle-like shape at quadratic cross-section [13] is more difficult to reconcile within a thermodynamic picture with the almost cubic Wulff shape shown in Fig. 2. In this case, the elongation cannot be rationalized with the c(2x2)-(100) reconstruction. If this reconstruction was stabilized through microfacetting of the exposed {110} lateral facets, this should also hold for the {100} facets themselves, invariably driving the particle cross-section to octagonal. As to date no other surface reconstruction of RuO$_2$(110) itself has been reported under oxygen-rich conditions, it is thus difficult to ascribe the deviation from the constrained Wulff shape to a lowered surface free energy of the lateral facets as compared to the (1x1)-terminations considered. Simultaneously, any missed reconstruction of apical facets would not help either, as it would only tend to further reduce the share of the lateral facets. At the given quadratic cross-section, this share is entirely determined by the ratio of $\gamma(110)/\gamma(111)$, which would need to be significantly lower to arrive at an elongated columnar shape. Considering its well-documented success in describing oxidation reactions at $RuO_2$ surfaces [16-18], we find it hard to assign such a large error, in particular in relative surface free energies, to shortcomings of the employed semi-local DFT functional. Instead, in our view a more likely explanation for the shape difference than unconsidered surface reconstructions modifying the Wulff shape within a thermodynamic picture is that the Narkhede crystals had not achieved their true equilibrium shape, but are dominated by kinetic limitations of the growth process. Also supported by possibly interesting catalytic function as discussed in the following, this motivates further experiments revisiting the shape-pretreatment relationship in particular at low calcination temperatures. One possibility to overcome kinetic limitations at such temperatures could thereby be to apply overpressures, which would allow to reach similarly high oxygen chemical potentials but at higher temperatures. This is illustrated in Fig. 2, where we show an additional temperature scale for an $O_2$ pressure of 10 atm.

The obtained constrained Wulff particle shapes and the discussed understanding of deviations in terms of surface reconstructions and kinetic limitations finally enables an interesting discussion of the catalytic activities reported in the powder catalyst studies [13-15]. Considering the established inactivity of c(2x2) reconstructed $RuO_2(100)$ towards CO oxidation [31], the initial inactivity of the as prepared Rosenthal crystals in oxidizing feed supports the view of a corresponding restructuring of all lateral facets already during the redox pretreatment. In contrast, the immediate activity of the Narkhede crystals in such a feed is well rationalized by the initial presence of the highly active O-rich (1x1)-(110) termination at the lateral facets. As discussed by Aßmann et al. [31], the slow degradation in both stoichiometric and oxidizing feed would then be consistent with the gradual reconstruction of the {110} facets to c(2x2)-(100) microfacets, accompanied with a morphology change to the octagonal habit.

With the lateral facets deactivated in either case, Rosenthal et al. ascribed the observed long-term catalytic function under oxidizing conditions to the apical facets [15]. The long induction time in such feed then suggests that the exposed, as prepared {101} and {111} stoichiometric terminations are essentially inactive, and only slowly transform to a more active phase. One possible candidate could be the (1x2)-$RuO_2$(101) reconstruction, which Kim et al. found to be highly reactive [37]. However, this structure was achieved through high-temperature annealing in UHV and should thus rather correspond to an O-deficient composition. Intermediate temperature CO oxidation in net oxidizing feeds points instead rather towards the stabilization of O-enriched structures as compared to those of the high-temperature calcination pretreatment. In this respect it is intriguing to note that the super-O-rich $RuO_2$(111)-termination, cf. Fig. 1, features exactly the kind of singly-coordinated "cus" O atoms that have served as the central rationalization for the known high activity of the $RuO_2$(110) surface [16-18]. If this termination was indeed active, the long induction times of the Rosenthal experiments could possibly be assigned to the stabilization of such O-enriched structures together with an increasing share of {111} facets at particles evolving towards their equilibrium shape at higher chemical potentials, cf. Fig. 2.

In fact, such a view puts even more emphasis on the discrepancy between the obtained constrained Wulff shape and the crystal habit reported by Narkhede et al.: If the columnar Narkhede crystals with their large share of lateral facets were indeed an artifact of an incomplete crystal shape equilibration at the low calcination temperature, so would be the reported deactivation through formation of the inactive c(2x2)-(100) phase at these lateral facets. The equilibrium shape obtained within the constrained Wulff construction at (0.1atm, 573K) instead barely features lateral facets at all. Actually, at only slightly lowered calcination temperatures (or slightly increased oxygen pressures), these facets become completely suppressed and the Wulff particles (apart from the small {001} apical cap) feature exclusively {111} apical facets. As such, it would be very interesting to revisit the crystal morphology after high chemical potential calcination and the concomitant catalytic activity in oxidizing feed. If the first-principles constrained Wulff shape was validated and the O-enriched {111} facets were indeed active as suggested by Rosenthal *et al.* [15], such pretreatment would result in catalyst particles that would not suffer from deactivation through the formation of the c(2x2)-(100) phase. In addition to the core-shell concept brought forward by Aßmann *et al.* [31], this could thus establish a second type of active $RuO_2$ catalyst with long-term stability in oxidizing feed.

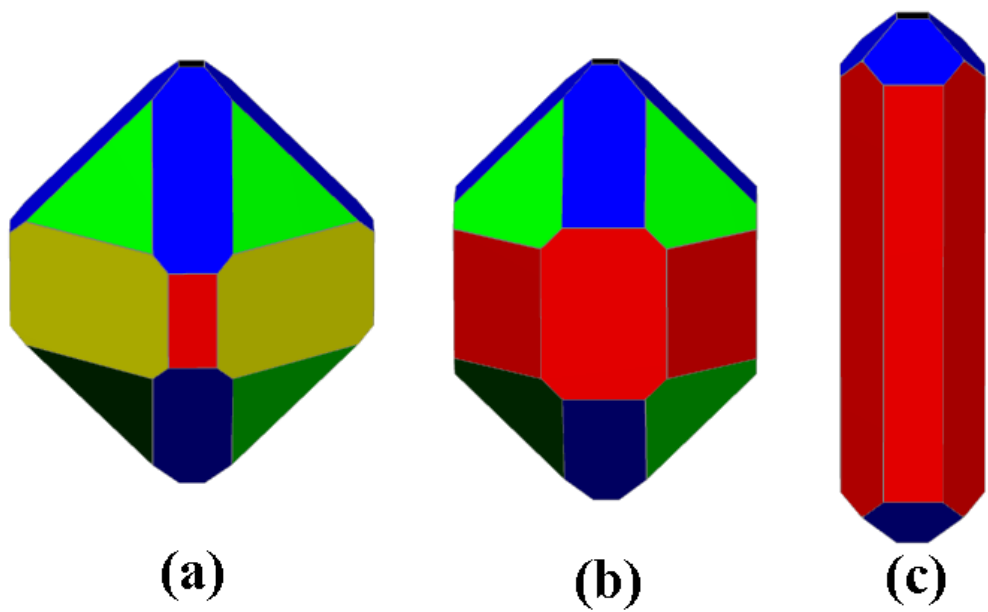

Figure 3: RuO2 crystal shape at $\Delta\mu_O$ = -1.24 eV corresponding to the calcination procedure (1atm, 1073K) employed in the Rosenthal experiments [13-14]: (a) First-principles constrained Wulff shape, cf. Fig. 2. (b) Shape resulting when reducing the surface free energy of all lateral facets to the computed value for γ(1x1)-(110) as minimum reduction induced by a restructuring and microfacetting to c(2x2)-(100) (indicated by drawing all lateral facets in the color code of the (100) facet, cf. Fig. 2). (c) Shape resulting when further reducing the surface free energy of all lateral facets to the point where the (111) apical facets disappear (see text).

## Conclusion

We have employed an *ab initio* thermodynamics Wulff construction to obtain the equilibrium shape of $RuO_2$ particles in an oxygen atmosphere. The results indicate a sensitive dependence of the crystal habit on the O chemical potential, with different facets exposed and a change of cross-section from quadratic to irregular octagonal. Overall these findings are fully consistent with the different crystal shapes reported for different calcination pretreatments in recent powder catalyst studies [13-15]. In detail, the experimental particles are much more columnar than the first-principles Wulff shapes though. For the higher temperature calcination employed by Rosenthal *et al.* [14,15] this is likely rationalized by a microfacetting of the lateral facets not contained in the present model. For the lower temperature calcination employed by Narkhede *et al.* [13] this rather points towards an incomplete shape equilibration in the pretreatment.

Correlating the reported initial activities towards CO oxidation in oxidizing feed with this detailed insight about the crystal habit supports the interpretation of Rosenthal *et al.* in terms of a prominent role of the hitherto little studied {101} and {111} apical facets. In particular, we argue that the deactivation reported by Narkhede *et al.* could be an artefact of the incomplete particle shape equilibration reached in the pretreatment. This strongly suggests revisiting the morphology-activity relationship for high chemical potential calcination. If the first-principles Wulff shape is validated, this could yield an interesting new type of active $RuO_2$ catalyst with long-term stability in oxidizing feed.

## Experimental Section

With the approximations discussed in the text, the working equation to evaluate the surface free energies entering the Wulff construction reads [7,8]:

$$\gamma(\Delta\mu_O) = \frac{1}{2A}\left\{E_{\text{slab}} - N_{\text{Ru}} E_{\text{RuO}_2(\text{bulk})} + (2N_{\text{Ru}} - N_{\text{O}})\left[\tfrac{1}{2} E_{O_2} + \Delta\mu_O\right]\right\},$$

where the factor ½ in front derives from the two sides of the symmetric slabs employed to describe the surface system. The required DFT input thus concerns the total energies of the surface system $E_{\text{slab}}$, of $RuO_2$ bulk $E_{\text{RuO2(bulk)}}$, and of an isolated $O_2$ molecule $E_{O2}$. All these DFT calculations were performed within the Cambridge Ab-Initio Simulation Total Energy Program (CASTEP) [38], treating electronic exchange and correlation with the generalized gradient approximation (GGA) functional due to Perdew, Burke and Ernzerhof (PBE) [39]. The core electrons were described by Vanderbilt ultrasoft library pseudo potentials, while the valence electrons were treated with a plane wave basis set with a cutoff energy of 450 eV. Supercell geometries with symmetric slabs and a vacuum separation exceeding 20 Å were employed to describe the different surface orientations and (1x1)-terminations. Specifically, we used 5 to 7 layers with (7x10x1), (7x7x1), (10x5x1), (6x7x1) and (6x6x1) Monkhorst-Pack reciprocal space sampling for $RuO_2(100)$, $RuO_2(001)$, $RuO_2(110)$, $RuO_2(101)$ and $RuO_2(111)$, respectively. For each orientation $E_{RuO2(bulk)}$ was computed at compatible k-meshes to allow for maximum error cancelation. Keeping only the middle slab layer fixed, all surfaces were fully relaxed until residual forces fell below 0.05 eV/Å. $E_{O2}$ was computed spin-polarized, in a (11Å x 12Å x 13Å) supercell with Gamma-point sampling. Systematic tests showed the computed surface free energies to be converged within 5 meV/Å$^2$ at the employed settings, which is sufficient for all conclusions made. The Wulff shape figures were created with WinXmorph [40].

## Acknowledgements

We gratefully acknowledge insightful discussions with Sebastian Matera.